\documentclass[12pt,a4paper]{article}

\usepackage{graphicx}
\usepackage{amssymb,amsfonts,amsmath}
\usepackage{cite}
\newcommand{\SC}{_{\mbox{\tiny C}}}
\newcommand{\SD}{_{\mbox{\tiny D}}}

\newcommand{\EQ}{Eq.~}
\newcommand{\EQS}{Eqs.~}
\newcommand{\FIG}{Fig.~}
\newcommand{\FIGS}{Figs.~}

\title{Evolution of cooperation driven by zealots}
\bigskip
\author{Naoki Masuda${}^{1*}$
\ \\
\ \\
${}^{1}$ 
Department of Mathematical Informatics,\\
The University of Tokyo,\\
7-3-1 Hongo, Bunkyo, Tokyo 113-8656, Japan
\ \\
$^*$ Corresponding author (masuda@mist.i.u-tokyo.ac.jp)}

\begin{document}
\setlength{\baselineskip}{24pt}

\maketitle

\newpage

\section*{Abstract}

Recent experimental results with humans involved in social dilemma games suggest that cooperation may be a contagious phenomenon and that the selection pressure operating on evolutionary dynamics (i.e., mimicry) is relatively weak. I propose an evolutionary dynamics model that links these experimental findings and evolution of cooperation. By assuming a small fraction of (imperfect) zealous cooperators, I show that a large fraction of cooperation emerges in evolutionary dynamics of social dilemma games. Even if defection is more lucrative than cooperation for most individuals, they often mimic cooperation of fellows unless the selection pressure is very strong. Then, zealous cooperators can transform the population to be even fully cooperative under standard evolutionary dynamics.

\newpage

Humans often behave cooperatively in social dilemma situations in which
withholding cooperative behavior is logically better.
In fact, cooperation in social dilemma games has been explained by
various mechanisms including kin selection, assortative interactions, group
competition, direct reciprocity (i.e., repeated interactions),
reputation-based indirect reciprocity, and spatial or network structure of
populations \cite{Nowak2006Science,Nowak2006book}.

In contrast, recent evidence suggests that cooperation at a population
level may occur as cascading in a social network
\cite{Fowler2010PNAS}.  Such a contagious view of cooperation is
distinct from the mechanisms governing cooperation explained above,
which assume that individuals at least try to maximize material
payoffs, a central assumption in game theory including evolutionary
game theory. In contagion, individuals may imitate others' behavior
without much caring the material payoff to the self and others.

The comtemporary results of the so-called
upstream reciprocity (also
called generalized exchange and pay-it-forward reciprocity; it is a type of indirect reciprocity) are
also consistent with contagious scenarios of cooperation.
Upstream reciprocity, in which
an individual helped by
somebody helps somebody else, is widely observed in
humans
\cite{Yamagishi1993SPQ,Molm2007AJS,Dufwenberg2001HomoOeco,Greiner2005JEconPsy,Stanca2009JEconPsy}. However,
theory assuming payoff maximization does not support upstream reciprocity
on its own \cite{Boyd1989SocN,Masuda2011PlosOne}. Cooperation on the basis of
upstream reciprocity is stable only in combination
with a different mechanism such as direct reciprocity
\cite{Nowak2007RoyalB}, mobility of players
across groups \cite{Hamilton2005RoyalB}, assortative interaction
\cite{Rankin2009Evol}, 
sufficiently frequent ingroup interaction \cite{Barta2011RoyalB},
and network reciprocity
\cite{Nowak2007RoyalB,Iwagami2010JTB,vanDoorn2012Evol}.

Contagion of suboptimal behavior or attitudes
seems to be even more common outside
social dilemma games. Even if individuals do not
like the behavior or social norm
(e.g., binge drinking in colleges) and are not forced to adopt it,
they often obey others.
In sociology, such a phenomenon
is interpreted under the framework of herd
behavior, pluralistic ignorance, and false enforcement
\cite{Centola2005AJS,Willer2009AJS}.

Contagion implies that individuals change behavior under
null or weak selection pressure. 
Consistent with this,
the selection pressure for humans playing the prisoner's dilemma game
was recently shown to be relatively weak;
subjects did not take the
imitate-the-best behavior with a probability of $\approx$ 30\%
\cite{Traulsen2010PNAS} (also see
\cite{Kirchkamp2007GEB}).

Nevertheless, even under weak selection pressure, unconditional
defection remains the unique Nash equilibrium of the social dilemma game.
Then, how can we explain cascades of cooperation found in experiments?

In mimicry-guided opinion
formation models in which the two competing opinions are equally strong,
a small number of zealot voters can
attract nonzealous players to the preferred opinion of the zealot
\cite{Mobilia2003PRL,Mobilia2007JSM,GalamJacobs2007PhysicaA,XieSreenivasan2011PRE,SinghSreenivasan2012PRE}. Motivated by these studies,
I show that a small fraction of
zealous cooperators can reliably induce cooperation at a population level.
It should be noted that opinion formation models and (social dilemma) games are fundamentally different in that only the latter involves strategic interactions and natural selection.
Examples of situations in which zealous cooperators in social dilemmas are witnessed
include military services and team sports \cite{Coleman1988SocTh}, and perhaps
charity campaigns. 
In the proposed mechanism, weak selection promotes cooperation. However,
the selection pressure does not have to be 
as weak as assumed in other theories of cooperation
(e.g., \cite{Nowak2004Nature,Ohtsuki2006Nature}).
The proposed mechanism does not require additional model components such as
the conformity bias \cite{Boyd1985book,Szolnoki2012SciRep}
or so-called cooperation facilitators that increment payoffs to cooperators, but not to defectors \cite{Mobilia2012PRE}.

\section*{Results}

I analyze the evolutionary dynamics given by
\EQS\eqref{eq:dx_c/dt} and \eqref{eq:dy_c/dt}.
I mainly examine the prisoner's dilemma game
described by a standard payoff matrix
given by
$R=1$, $T>1$, and $S=P=0$. 
When zealous players are absent (i.e., $y=0$), D is the only Nash equilibrium
(i.e., $x\SC=0$).
I explore the possibility that
cooperation is stabilized among ordinary players in the presence of zealous
players.

\subsection*{Prisoners' dilemma with perfect zealous cooperators}

In this section, I consider the case in which the zealous players
always cooperate (i.e., $p=1$, $y\SC = y$). In this case,
\EQ\eqref{eq:dx_c/dt} is reduced to
\begin{equation}
\frac{{\rm d}x\SC}{{\rm d}t} = \frac{1-x\SC}{\left<\pi\right>}
\left\{ (1-w) y + w\frac{(x\SC+y)(x\SC+y-Tx\SC)}{1+y}\right\}.
\end{equation}
Because the coefficient of $x\SC^2$ in
$f(x\SC)\equiv (1-w) y + w(x\SC+y)(x\SC+y-Tx\SC)/(1+y)$
is negative and $f(0)=(1-w)y+wy^2/(1+y)>0$, the dynamics has at most one internal equilibrium, which is stable if it exists. When $f(1)>0$, i.e.,
\begin{equation}
T\le 1+\frac{y}{w},
\label{eq:phase transition p=1}
\end{equation}
$dx\SC/dt>0$ ($0\le x\SC<1$) holds true such that
the only equilibrium is located at $x\SC^*=1$. Then, all the
players eventually cooperate. Equation~\eqref{eq:phase transition p=1}
indicates that 
weak selection and the presence of many zealous players
facilitate such full cooperation.
If $T>1+y/w$, the stable equilibrium $x\SC^*$ is given by
\begin{equation}
x\SC^* = \frac{-wy(T-2)+\sqrt{w^2y^2(T-2)^2+4w(T-1)y(y+1-w)}}{2w(T-1)}.
\end{equation}
It should be noted that $0<x\SC^*<1$.

\begin{figure}[h]
\begin{center}
\includegraphics[height=6cm]{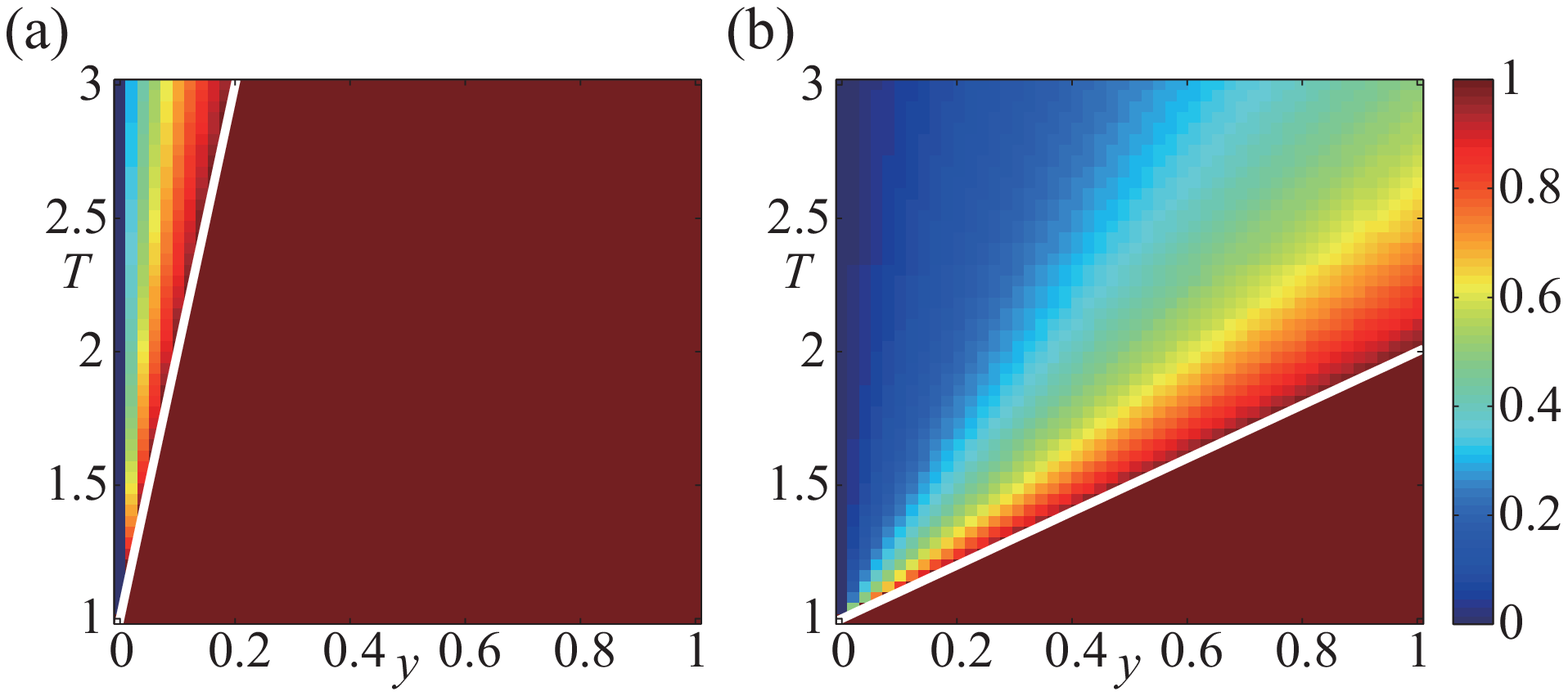}
\caption{Fraction of cooperators among the ordinary players in the presence of perfectly zealous cooperators (i.e., $p=1$). The lines represent $T=1+y/w$. I used a typical payoff matrix of the prisoner's dilemma game given by $R=1$, $T>1$, and $S=P=0$. I set (a) $w=0.1$ and (b) $w=1$.}
\label{fig:xc* p=1}
\end{center}
\end{figure}

The equilibrium fraction of cooperators among the ordinary players
is shown as a function of $y$ and $T$ in the case of
relatively weak ($w=0.1$) and strong
($w=1$) selection in \FIGS\ref{fig:xc* p=1}(a)
and \ref{fig:xc* p=1}(b), respectively. The lines represent
$T=1+y/w$ and separate the full cooperation phase and
the partial cooperation phase.
Figure~\ref{fig:xc* p=1} indicates that
the fraction of cooperators 
is mainly determined by $y/w$ and is larger with
the weak selection than the strong selection.

\subsection*{Prisoner's dilemma with imperfect zealous cooperators}

In fact, zealous players may not perfectly cooperate. Therefore, I
investigate
evolutionary dynamics given by \EQS\eqref{eq:dx_c/dt} and
\eqref{eq:dy_c/dt} with
$p<1$ by numerically integrating them.
I use the Euler-Maruyama integration scheme with $dt=0.01$.
For a fixed set of parameter values, I
started the evolutionary dynamics from various initial conditions,
i.e., $(x\SC, y\SC)=(0.05i, 0.05j)$, where $i$ and $j$ are integers
and
$1\le i, j\le 19$,
and confirmed that the equilibrium is independent of the initial
condition.

\begin{figure}[t]
\begin{center}
\includegraphics[height=12.5cm]{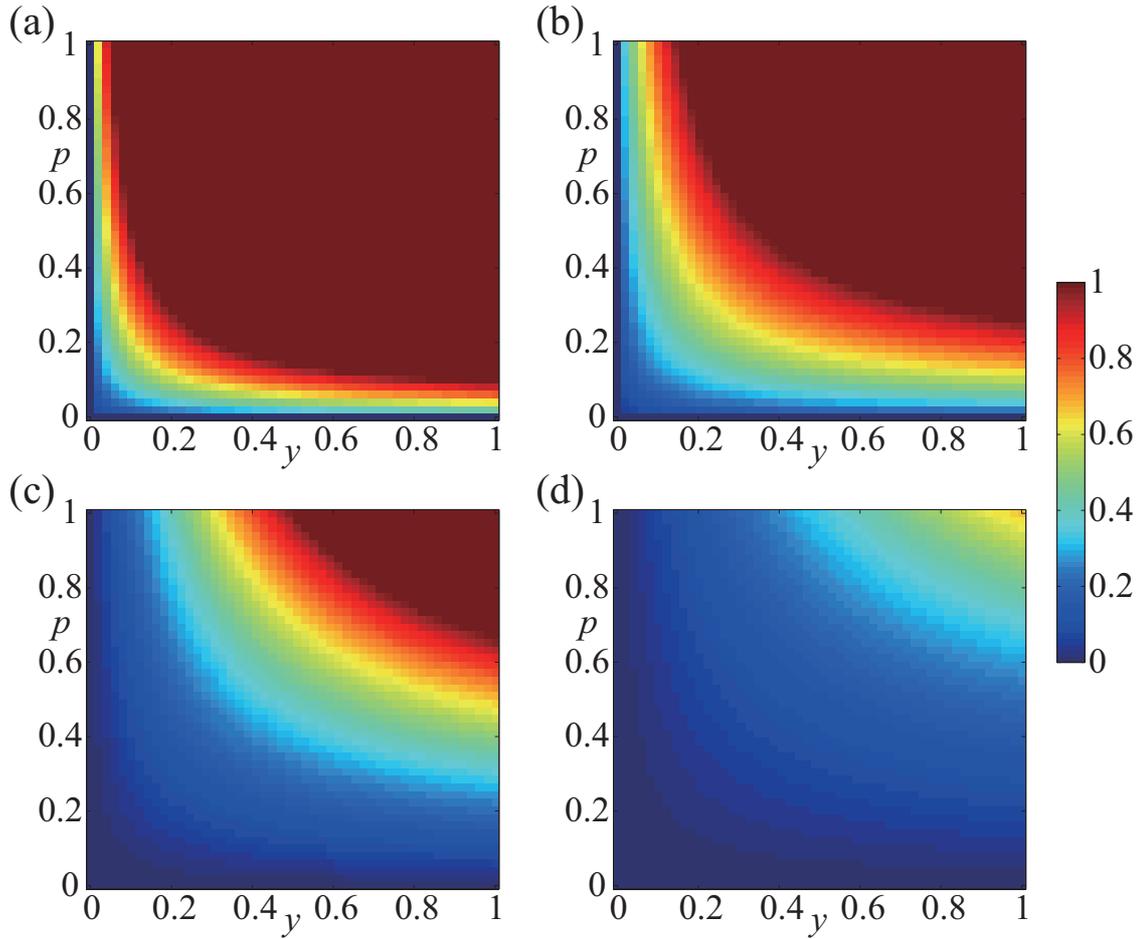}
\caption{Fraction of cooperators among the ordinary players as a function of the additional density of zealous players (i.e., $y$) and the probability of unconditional cooperation for zealous players (i.e., $p$). I set $R=1$ and $S=P=0$. (a) $w=0.1$, $T=1.5$. (b) $w=0.1$, $T=2.5$. (c) $w=1$, $T=1.5$. (d) $w=1$, $T=2.5$.}
\label{fig:vary y and p}
\end{center}
\end{figure}

The equilibrium fraction of cooperators among the ordinary players 
is shown for $(w, T)=(0.1, 1.5)$, (0.1, 2.5), (1, 1.5), and (1, 2.5) in \FIGS\ref{fig:vary y and p}(a)
\ref{fig:vary y and p}(b),
\ref{fig:vary y and p}(c),
\ref{fig:vary y and p}(d),
respectively.
 When $w=0.1$, nearly perfect cooperation is obtained unless
the fraction of zealous players (i.e., $y$) and the 
probability that zealous players unconditionally cooperate (i.e., $p$)
 are both small. This is also the case when the temptation
payoff $T$ is rather large (\FIG\ref{fig:vary y and p}(b); $T=2.5$).
Even when $w=1$, a considerable amount of
cooperation (e.g., 0.4) is observed for a wide parameter range when
$T=1.5$ (\FIG\ref{fig:vary y and p}(c)). 

\begin{figure}[h]
\begin{center}
\includegraphics[height=6cm]{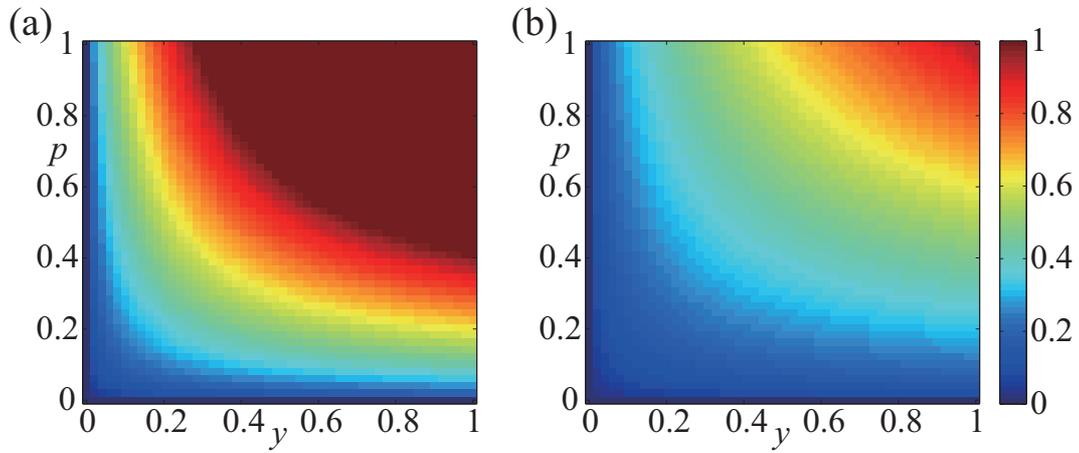}
\caption{Fraction of cooperators among the ordinary players in the prisoner's dilemma game when the pairwise comparison rule is used for the updating. I set $R=1$, $S=P=0$, and $\overline{\beta}=0.5$. (a) $T=1.5$. (b) $T=2.5$.}
\label{fig:Fermi}
\end{center}
\end{figure}

To confirm that the results are not specific to the Moran type of the reproduction process, I also implemented the
so-called pairwise comparison rule (Methods)
\cite{Blume1993GEB,SzaboToke1998pre,Nowak2004Nature,TraulsenNowakPacheco2006pre}.
In short, in the pairwise comparison rule, the probability of the
strategy replacement is a sigmoid function of the difference between
the fitness of two randomly selected players. This update rule is
implicated in recent laboratory experiments
\cite{Traulsen2010PNAS}. 
Numerical results for the prisoner's dilemma game with $T=1.5$ and $T=2.5$
under the pairwise comparison rule with $\overline{\beta}=0.5$ are shown in 
\FIGS\ref{fig:Fermi}(a) and \ref{fig:Fermi}(b), respectively. The results are qualitatively the same as those under the Moran process
(\FIG\ref{fig:vary y and p}).
In the rest of the present paper, I use the Moran process.

\begin{figure}[h]
\begin{center}
\includegraphics[height=6cm]{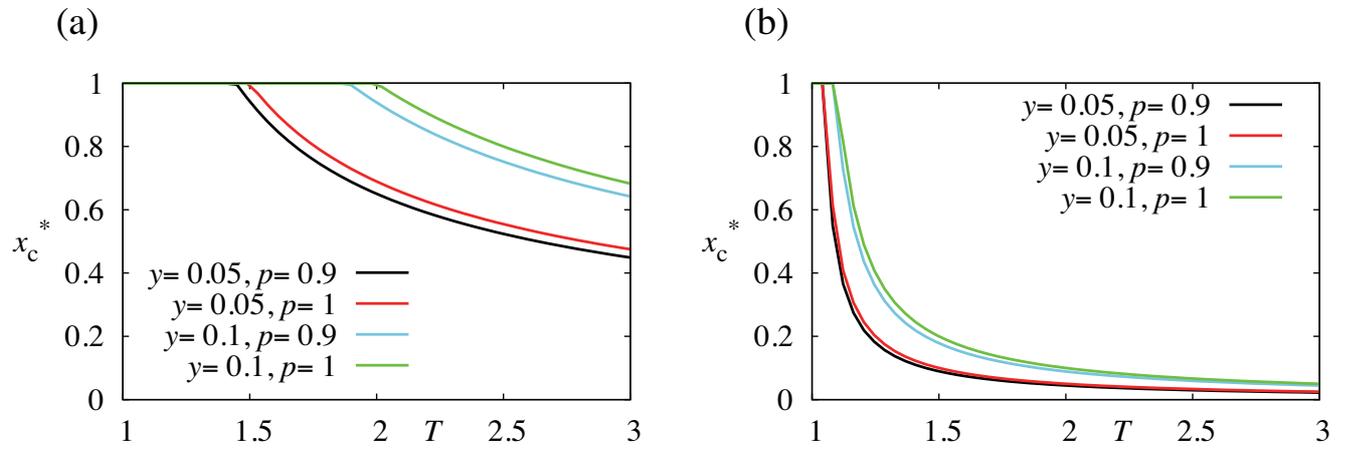}
\caption{Fraction of cooperators among the ordinary players as a function of the temptation payoff (i.e., $T$). I set $R=1$ and $S=P=0$. (a) $w=0.1$. (b) $w=1$.}
\label{fig:vary T}
\end{center}
\end{figure}

The fraction of cooperators among the ordinary players for various values of
$T$ is shown in \FIG\ref{fig:vary T}. I used
two large values of $p$ ($p=0.9$ and $p=1$) and two small values of $y$
($y=0.05$ and $y=0.1$).
Under both weak selection
(\FIG\ref{fig:vary T}(a); $w=0.1$) and strong selection
(\FIG\ref{fig:vary T}(b); $w=1$), the results do not depend much on
the value of $p$ for large $p$. This behavior is also evident in 
\FIG\ref{fig:vary y and p}.
Therefore, the theoretical results obtained in the previous section
for the case of perfectly cooperating zealous players (i.e., $p=1$)
are translated to the case of imperfect zealots (i.e., $p<1$)
without much change.
In contrast, the fraction of cooperation is sensitive to the
density of zealots (i.e., $y$).

\subsection*{Snowdrift game}

The emergence of cooperation owing to the combination of 
ordinary players and zealots
is not restricted to the case of
the prisoner's dilemma. In this section, I briefly examine
the snowdrift game, also known as the chicken game and the hawk-dove
game \cite{Maynardsmith1982book,Sugden1986book,Hauert2004Nature}.
The payoff matrix of a standard snowdrift game is given by
$R=\beta-0.5$, $T=\beta$, $S=\beta-1$, and $P=0$, where $\beta >1$
\cite{Hauert2004Nature}. I set $\beta=1.5$ such that
the stable fraction of C in the absence of zealots is
given by $(2\beta-2)/(2\beta-1) = 0.5$.

\begin{figure}[h]
\begin{center}
\includegraphics[height=6cm]{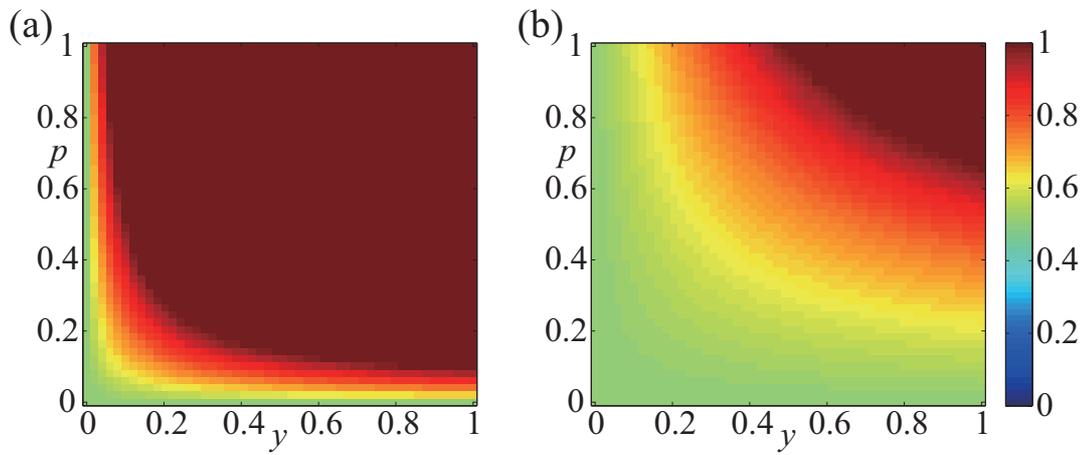}
\caption{Fraction of cooperators among the ordinary players as a function of $y$ and $p$ in the snowdrift game. I set $R=\beta-0.5$, $S=\beta-1$, $T=\beta$, $P=0$, and $\beta=0.5$. (a) $w=0.1$. (b) $w=1$.}
\label{fig:snowdrift}
\end{center}
\end{figure}

The fraction of cooperation among the ordinary
players is shown for $w=0.1$ and $w=1$
in \FIGS\ref{fig:snowdrift}(a) and \ref{fig:snowdrift}(b), respectively.
Figure~\ref{fig:snowdrift} indicates that the fraction of cooperators
is much larger than 0.5 in a wide parameter range,
particularly when $w=0.1$.

\section*{Discussion}

I showed that a small fraction of zealous cooperators can guide
cooperation of players that obey a standard evolutionary dynamics. The
numerical results indicate that the
zealous players do not have to be perfectly zealous cooperators. The proposed
mechanism operates better when the selection pressure is weak and
the density of zealous players is large.
Although I used the Moran process and the replicator dynamics,
the results do not qualitatively change if
a different strategy update rule called the pairwise comparison rule is used.

All the present results are independent of the
initial condition. Therefore, if the dynamics is initiated from a
small density of cooperation, perhaps only among zealots,
cooperation can cascade to prevail in the population. I emphasize that
the cascade can occur even if most players prefer defection to
cooperation to some extent. My results may provide theoretical
underpinning of cascades of cooperation
\cite{Fowler2010PNAS} and upstream reciprocity
\cite{Yamagishi1993SPQ,Molm2007AJS,Dufwenberg2001HomoOeco,Greiner2005JEconPsy,Stanca2009JEconPsy}
observed in human subjects.
In contrast, the emergence and maintenance of cooperation based on
conformity bias \cite{Boyd1985book} requires that a majority of players initially cooperates.

Crucial assumptions underlying the proposed explanation of cooperation
are the stochasticity
of the dynamics and weak selection
(i.e., small $w$). Weak selection is often employed in theoretical
studies because, among other things, Taylor expansion on $w$
often leads to analytical conditions for cooperation 
(e.g., \cite{Nowak2004Nature,Ohtsuki2006Nature}).
However, 
experiments with human subjects present evidence against
excessively weak selection \cite{Traulsen2010PNAS}.
I referred to the intensity of selection equal to $w=0.1$
as weak selection. This value of $w$ may not be too small to 
violate the reality.
In general, the effective
intensity of selection depends on
the payoff matrix and the strategy update rule
as well on the $w$ value. Nevertheless, the following simple calculus
may help: the largest and smallest possible fitness values
in the prisoner's dilemma
used in this study are equal to 
$1-w+wT$ and $1-w$, respectively. Therefore,
$w=0.1$ indicates that the ratio of the two fitness values for $T=2$,
for example, is
equal to $(1-w+wT)/(1-w)=11/9$. If this ratio should exceed 2,
$w>1/3$ is required under $T=2$. Although I only examined
the extreme two cases, i.e., $w=0.1$ and $w=1$, 
the results shown in the figures
imply that much cooperation will
be observed with $w=1/3$.
In fact, some cooperation is observed even with $w=1$ if proper
conditions are met (\FIGS\ref{fig:xc* p=1}(b) and \ref{fig:vary y and p}(c)).

It should be noted that I assumed that players, either ordinary or zealous, have the same strength of influence on others.
Although the heterogeneity in the influence of individuals, i.e., power,
would shape collective behavior of humans, the present contribution is not about the power but about the relationship between zealots, weak selection, contagion, and cooperation.

I did not ask the origin of zealous cooperators. Trivially, they will
not emerge as a result of evolution unless other games or dynamics are
simultaneously considered. One interpretation of this assumption 
is that zealous players are not
interested in maximizing the material payoff. Zealous cooperators are
found in some real situations \cite{Coleman1988SocTh}.
Another interpretation is that zealots are payoff maximizers but 
have different payoff functions. In theory of collective action, heterogeneity in interests and resources of individuals are suggested to elicit collective action to solve the free rider problem (see \cite{Oliver1993ARS} for a review). Although the present mechanism is independent of that of collective action, zealots may perceive payoffs differently from ordinary players such that cooperation may not incur social dilemma for zealots.

\section*{Methods}

\subsection*{Model}

I consider evolutionary dynamics of
an infinite well-mixed population in which each pair of players is involved in
the symmetric two-player two-strategy game once per generation.
The payoff matrix is defined by
\begin{equation}
\bordermatrix{
 & {\rm C} & {\rm D} \cr
{\rm C} & R & S \cr
{\rm D} & T & P \cr}, \;
\label{eq:payoff}
\end{equation}
where the entries of
\EQ\eqref{eq:payoff}
represent the payoffs that the row player gains. Each row (column)
corresponds to the action of the row (column) player, i.e.,
cooperation (C) or defection (D).

I assume two types of players. A player of the first type, called
ordinary player, obeys a standard
evolutionary dynamics described below. A player of the second type,
called zealous player, may obey the evolutionary dynamics or 
unconditionally cooperate.

The evolutionary dynamics is defined as follows. 
The summation of the payoff over all the opponents defines
the aggregated payoff to a player. The fitness, i.e., the propensity
to reproduce, of a player is a linear function of the payoff. The
proportionality constant controls the intensity of selection.
At the end of each generation, 
a single player whose strategy (i.e., C or D) is replaced
is selected with the equal
probability from the population.
If the selected player is ordinary player,
a parent, which is either ordinary or zealous player,
is selected from the entire population with the probability proportional to the
fitness.
Then, the strategy of the updated player is replaced by that
of the parent player.
This update process is equivalent to the Moran process, a standard
model of the birth-death process (e.g., \cite{Nowak2006book}).
If the updated player is zealous player,
its strategy turns to C with probability $p$. With
probability $1-p$, the updated player obeys the rule used by the
ordinary player to adopt
the strategy of a parent selected with
the probability proportional to the fitness.

I normalize the density of the ordinary players to unity and denote
the added density of zealous players by $y (\ge 0)$.
The densities of cooperators among the ordinary and zealous
players are denoted by $x\SC$ ($0\le x\SC\le 1$) and $y\SC$
($0\le y\SC\le y$), respectively.
The mean fitness, with the divisive factor $1+y$ (i.e., the total
population density) intentionally neglected, is defined by
\begin{equation}
\left<\pi\right>\equiv (x\SC+y\SC)\pi\SC + (1-x\SC+y-y\SC)\pi\SD,
\end{equation}
where
\begin{equation}
\pi\SC= 1-w + \frac{w\left[(x\SC+y\SC)R + (1-x\SC+y-y\SC)S\right]}{1+y}
\label{eq:pi c}
\end{equation}
and
\begin{equation}
\pi\SD= 1-w + \frac{w\left[(x\SC+y\SC)T + (1-x\SC+y-y\SC)P\right]}{1+y}
\label{eq:pi d}
\end{equation}
are the fitness to a C and D player, respectively, and $w$ ($0\le w\le 1$)
indicates the intensity of selection
\cite{Nowak2004Nature,Ohtsuki2006Nature}. Equations~\eqref{eq:pi c} and
\eqref{eq:pi d} indicate that
the payoff to a player per opponent is translated to
the fitness with proportionality constant $w$.
I assume $\left<\pi\right> > 0$ such that the selection of the parent
player with the probability proportional to the fitness is well defined.

In the reproduction phase, a cooperator and defector are selected as
parent with probability $(x\SC+y\SC)\pi\SC/\left<\pi\right>$ and
$(1-x\SC+y-y\SC)\pi\SD/\left<\pi\right>$, respectively. 
Therefore, the dynamics of the fraction of cooperators among the ordinary players is given by
\begin{equation}
\frac{{\rm d}x\SC}{{\rm d}t} =
\frac{1}{\left<\pi\right>}\left[(x\SC+y\SC)\pi\SC(1-x\SC) -
  (1-x\SC+y-y\SC)\pi\SD x\SC \right].
\label{eq:dx_c/dt}
\end{equation}
When $y=0$, \EQ\eqref{eq:dx_c/dt} is equivalent to the meanfield
equation of the Moran process.
If the divisive factor $\left<\pi\right>$, which just controls the
time scale of the dynamics in this special case, is neglected, 
\EQ\eqref{eq:dx_c/dt} is reduced to the usual replicator dynamics.

The dynamics of the density of cooperators among the zealous players is
given by
\begin{equation}
\frac{{\rm d}y\SC}{{\rm d}t} = p(y-y\SC) + \frac{1-p}{\left<\pi\right>}
\left[(x\SC+y\SC)\pi\SC(y-y\SC) - (1-x\SC+y-y\SC)
\pi\SD y\SC \right].
\label{eq:dy_c/dt}
\end{equation}
When $p=1$, zealous players always cooperate (i.e., $y\SC = y$). In this case,
\EQ\eqref{eq:dx_c/dt}, with $\left<\pi\right>$ in the denominator neglected,
is equivalent to Example 2 given in
\cite{Chatterjee2012JTB}.

\subsection*{Pairwise comparison rule}

In the so-called pairwise
comparison rule, the probability that the replacement occurs
depends on the difference between the payoffs to two randomly selected players. At
the end of each generation, I randomly select two players from the population
without bias. If the two players are both cooperators or both
defectors, nothing takes place. Otherwise, C replaces D with
probability $1/\left[1+e^{-\overline{\beta}(\pi\SC-\pi\SD)}
\right]$, and D replaces C with probability $1-
1/\left[1+e^{-\overline{\beta}(\pi\SC-\pi\SD)}
\right] = 
1/\left[1+e^{-\overline{\beta}(\pi\SD-\pi\SC)}
\right]$
\cite{Blume1993GEB,SzaboToke1998pre,Nowak2004Nature,TraulsenNowakPacheco2006pre}.
The intensity of selection is controlled by $\overline{\beta} (\ge 0)$.

The evolutionary dynamics for the infinite population
under the pairwise comparison rule is represented by
\begin{equation}
\frac{{\rm d}x\SC}{{\rm d}t}=\frac{2}{1+y}\left[
\frac{(1-x\SC)(x\SC+y\SC)}{1+e^{-\overline{\beta}(\pi\SC-\pi\SD)}} 
- \frac{x\SC(1-x\SC+y-y\SC)}{1+e^{-\overline{\beta}(\pi\SD-\pi\SC)}}
\right]
\end{equation}
and
\begin{equation}
\frac{{\rm d}y\SC}{{\rm d}t}=p(y-y\SC) + \frac{2(1-p)}{1+y}\left[
\frac{(y-y\SC)(x\SC+y\SC)}{1+e^{-\overline{\beta}(\pi\SC-\pi\SD)}} 
- \frac{y\SC(1-x\SC+y-y\SC)}{1+e^{-\overline{\beta}(\pi\SD-\pi\SC)}}
\right].
\end{equation}


\section*{Acknowledgments}

I thank Mitsuhiro Nakamura and Shoma Tanabe for careful reading of the manuscript and Masanori Takezawa for valuable discussions. This work
is supported by Grants-in-Aid for Scientific Research (Nos. 20760258 and 23681033, and Innovative Areas ``Systems Molecular Ethology''(No. 20115009)) from MEXT, Japan. 

\section*{Additional information}

\textbf{Competing financial interests:} The author declares no competing financial interests.

\newpage
\clearpage

\clearpage

\clearpage

\clearpage

\clearpage

\end{document}